\documentclass[12pt]{article}
\pdfoutput=1
\usepackage{putex}
\usepackage{graphicx}
\usepackage{epstopdf}
\usepackage{enumerate}
\usepackage{cite}
\usepackage{tensor}
\usepackage{slashed}

\newcommand{\beq}{\begin{equation}}
\newcommand{\eeq}{\end{equation}}
\newcommand{\ba}{\begin{array}{ccc}}
\newcommand{\ea}{\end{array}}
\newcommand{\nn}{\nonumber \\}

\def\bea{\begin{eqnarray}}
\def\eea{\end{eqnarray}}

\newcommand{\CP}{\mathbb{CP}}

\newcommand{\R}{\mathbb{R}}

\newcommand{\es}[2] {\begin{equation} \label{#1} \begin{split} #2 \end{split} \end{equation}}

\def\mop#1{\mathop{\rm #1}\nolimits}

\def\coth{\mop{coth}}

\newcommand{\abs}[1]{\left\lvert #1 \right\rvert}

\usepackage{setspace} %line spacing
\begin{document}

\preprint{MIT-CTP-4447}

\institution{MITCTP}{Center for Theoretical Physics, Massachusetts Institute of Technology, Cambridge, MA 02139, USA}
\institution{Harvard}{Department of Physics, Harvard University, Cambridge, MA 02138, USA}

\title{Monopoles in $2+1$-dimensional conformal field theories with global $U(1)$ symmetry}

\authors{Silviu S.~Pufu\worksat{\MITCTP} and Subir Sachdev\worksat{\Harvard}}

\abstract{
In $2+1$-dimensional conformal field theories with a global $U(1)$ symmetry, monopoles can be introduced through a background gauge field that couples to the $U(1)$ conserved current.  We use the state-operator correspondence to calculate scaling dimensions of such monopoles.  We obtain the next-to-leading term in the $1/N_b$ expansion of the Wilson-Fisher fixed point in the theory of $N_b$ complex bosons.
 }

\date{March 2013}

\maketitle

%\tableofcontents

%%%%%%%%%%%%%%%%%%%%%%%%%%%%%%%%%%%%%%%%%%%%%%%%%%%%%%%%%
\section{Introduction}
\label{sec:intro}
%%%%%%%%%%%%%%%%%%%%%%%%%%%%%%%%%%%%%%%%%%%%%%%%%%%%%%%%%

Polyakov \cite{polyakov} introduced monopoles in 2+1 dimensions  as instanton tunneling events in compact gauge theories. 
The proliferation of these monopoles leads to confinement and to the absence of a Coulomb phase in such gauge theories,
provided there are no gapless matter fields that can suppress the monopoles. 
In condensed matter physics, two-dimensional lattice quantum antiferromagnets can be written as compact $U(1)$ gauge theories at 
strong coupling \cite{ba}: 
here, monopole events are accompanied by Berry phases \cite{haldane,rsl}, which are
responsible for valence bond solid order in the confining phase \cite{rsl,rsb}.

We can also consider monopole operators at conformal fixed points of $2+1$-dimensional gauge theories \cite{murthy,kapustin2,kapustin3,senthil1,senthil2,hermelemono,hermeleo4,benna}. These are gauge-invariant primary
operators that determine important aspects of the structure of the conformal field theory (CFT). 
In the application to antiferromagnets, the scaling dimension of the monopole operator determines the power-law decay of the valence bond solid order at ``deconfined'' quantum critical points \cite{senthil1,senthil2,ribhu,kedar}. 

This paper will consider a different class of monopoles in 2+1 dimensions. We consider CFTs with a {\em global\/} $U(1)$
symmetry. The CFT may also have fluctuating gauge fields, but these play no role in the construction of such monopoles. 
Instead, the monopole is introduced by a {\em background\/} $U(1)$ gauge field that couples to the CFT conserved current.  %by gauging the global $U(1)$ symmetry.  
A monopole with charge $q$ inserted at $r = r_0$, which we henceforth denote by ${\cal M}_q(r_0)$, corresponds to a background gauge field configuration whose field strength\footnote{In standard vector notation, instead of $f$ we would use the magnetic field $\beta = *f$, which can be also written as $\vec{\beta} = \vec{\nabla}  \times \vec{\alpha}$.  Eq.~\eqref{qUnits} becomes $\int_{S^2} \vec{\beta} \cdot d\vec{A} = 2 \pi q$, where $d\vec{A}$ is the oriented area element.}
 \es{flux}{
  f_{\mu\nu} = \partial_\mu \alpha_\nu - \partial_\nu \alpha_\mu 
 }
integrates to $2 \pi q$ over any small two-sphere surrounding the insertion point:
 \es{qUnits}{
  \int_{S^2} f = 2 \pi q \,.
 }
As we will see explicitly in Section~\ref{sec:method}, each such background monopole comes associated with a Dirac string that starts at $r=r_0$.  If the matter fields have integer $U(1)$ charges, the Dirac string is not observable provided that $q$ is an integer.

Such monopoles appear to not have been considered until recently \cite{willett,dopedcft}.  They do not correspond to operators in the CFT in a strict sense;  instead, they should be rather thought of as non-local background sources to which we couple our CFT\@.  Studying the response of the CFT to such background sources provides useful information about the CFT, which can be used, for instance, to test various dualities \cite{willett}.   In addition, these monopole insertions have been argued to play a crucial role in the structure of the compressible quantum phases that are obtained when a non-zero chemical potential is applied to the global $U(1)$ charge \cite{fi,dopedcft}. Specifically, they serve to quantize the $U(1)$ charge and to determine the lattice spacing of Wigner crystal states such that there are an integer number of particles per unit cell; they are also important in determining the period of Friedel oscillations \cite{fi,eva} of compressible states that do not break translational symmetries and may have ``hidden'' Fermi surfaces \cite{dopedcft,liza,tadashi1,hyper}.  In this paper we will restrict our attention to CFTs to which no external chemical potential has been applied.

The definition of ${\cal M}_q$ presented above is imprecise, partly because the condition \eqref{qUnits} does not specify $f$ uniquely, and partly because we have not specified the allowed behavior of the charged matter fields close to the singularity at $r = r_0$.  Just as in the case of monopole operators in gauge theories \cite{kapustin2},  a precise definition can be given through the state-operator correspondence, or, more precisely, through an extension thereof to the present case.  According to the state-operator correspondence, any local operator of a CFT inserted at the origin of $\R^3$ corresponds to a normalizable state of the CFT on $S^2 \times \R$, where the $\R$ coordinate is interpreted as Euclidean time.  A monopole insertion ${\cal M}_q$ is by no means a local operator, but it can nevertheless be defined as corresponding to the vacuum on $S^2$ (as opposed to any other excited state) in the presence of $q$ units of background magnetic flux (as in \eqref{qUnits}) that is uniformly distributed throughout the $S^2$.  

The monopole insertion defined above is a Lorentz scalar.  It also has a well-defined scaling dimension $\Delta_q$ in the following sense.  If we consider a background gauge field configuration $\alpha_\mu$ corresponding to a monopole of strength $q$ at $r = r_1$ and one of strength $-q$ at $r = r_2$, the partition function in the presence of these two monopole insertions has power-law decay with the relative distance $\abs{r_1 - r_2}$, namely
 \es{TwoPoint}{
  \langle {\cal M}_q(r_1) {\cal M}_{-q}(r_2) \rangle = \frac{ \int \mathcal{D} \phi \exp \left( - \int d^3 x \mathcal{L} [\alpha] \right)}{\int \mathcal{D} \phi \exp \left( - \int d^3 x \mathcal{L}  \right)}  \propto \frac{1}{\abs{r_1 - r_2}^{2 \Delta_q}} \,,
 } 
where we denoted by ${\cal L}[\alpha]$ the Lagrangian of the CFT coupled to $\alpha_\mu$.  One can extract $\Delta_q$ from the exponent in \eqref{TwoPoint}.  Equivalently, in view of the definition of ${\cal M}_q$ through the state-operator correspondence described above, one can also map a single monopole insertion on $\R^3$ to $S^2 \times \R$ and identify $\Delta_q$ with the ground state energy on $S^2$ in the presence of $q$ units of background magnetic flux.  Explicitly, we have
 \es{DeltaFormula}{
  \Delta_q = {\cal F}_q \equiv - \log Z_q \,,
 } 
where $Z_q$ is the partition function on $S^2 \times \R$, and ${\cal F}_q$ the corresponding free energy.

The question that we will address in this paper concerns the scaling dimensions $\Delta_q$ of the monopole insertions ${\cal M}_q$ in simple CFTs.  To calculate these scaling dimensions, we will use \eqref{DeltaFormula}.  For certain free CFTs with global $U(1)$ symmetry, one can infer $\Delta_q$ from existing results in the literature.  A simple example is the free CFT of $N_f$ complex fermions.  The Lagrangian
 \es{FermionTheory}{
  {\cal L}_f = \sum_{a = 1}^{N_f}  \psi_a^\dagger  (i  \slashed{\partial} ) \psi_a 
 }
is invariant under a $U(N_f)$ global symmetry under which $\psi_a$ transforms as a fundamental vector with $N_f$ components.  We can consider the diagonal $U(1)$ subgroup of $U(N_f)$, which we couple to a $U(1)$ background gauge field such that the modified Lagrangian is
 \es{FermionGauge}{
  {\cal L}_f[\alpha] = \sum_{a = 1}^{N_f}  \psi_a^\dagger  (i  \slashed{\partial} + \slashed{\alpha} ) \psi_a \,.
 }
As above, we consider monopole insertions of $q$ units of background magnetic flux.  Using \eqref{DeltaFormula}, $\Delta_q$ can be computed from the partition function on $S^2 \times \R$, which is now a Gaussian integral because the Lagrangian \eqref{FermionGauge} is quadratic in $\psi_a$ and there are no interactions.  The same Gaussian integral was calculated in Ref.~\cite{kapustin2} as part of a slightly different problem:  The authors of Ref.~\cite{kapustin2} were interested in computing the scaling dimensions of monopole operators in three-dimensional QED with $N_f$ flavors, which is the same theory as \eqref{FermionGauge}, with the exception that the gauge field $\alpha_\mu$ would be dynamical.  While the leading large $N_f$ result of Ref.~\cite{kapustin2} is only approximate for QED (because there are corrections coming from the fluctuations of the gauge field), in the free fermion theory \eqref{FermionGauge} one obtains an exact result that holds at all $N_f$.\footnote{We should restrict to $N_f$ even in order to avoid a parity anomaly.}   We reproduce the dimensions $\Delta_q$ for the first few lowest values of $q$ in Table~\ref{FermionMonopoles}.
 \begin{table}[ht!]
%\caption{default}
\begin{center}
\begin{tabular}{c|c}
  $q$ & $\Delta_q / N_f$ \\
  \hline \hline
  $0$ & $0$ \\
  $1$ & $0.265$ \\
  $2$ & $0.673$ \\
  $3$ & $1.186$ \\
  $4$ & $1.786$ \\
  $5$ & $2.462$
\end{tabular}
\end{center}
\caption{The scaling dimensions of the monopole insertions ${\cal M}_q$ in the free theory of $N_f$ fermions corresponding to the diagonal $U(1)$ subgroup of the global $U(N_f)$ symmetry group.  These results are exact in $N_f$.}
\label{FermionMonopoles}
\end{table}%
Similar results for a non-supersymmetric free theory of $N_b$ complex scalars are presented in Appendix~\ref{SCALAR}.

In this paper we are interested in the more complicated case of an {\em interacting} CFT with global $U(1)$ symmetry.  The simplest such CFT is the $XY$ model, described by the Wilson-Fisher fixed point of the $\phi^4$ field theory of a complex scalar field $\phi$.  Starting with the Lagrangian
 \es{lxy}{
   \mathcal{L}_{XY} = |\partial_\mu \phi|^2 + s |\phi|^2 + u |\phi|^4 \,,
 }
the Wilson-Fisher fixed point is reached in the infrared provided that the coefficient $s$ is tuned to zero.  This theory has a global $U(1)$ symmetry under which $\phi$ is rotated by a phase, and it is this $U(1)$ symmetry that we couple to a background gauge field $\alpha_\mu$.  The Lagrangian in the presence of $\alpha_\mu$ is
 \es{LagBackground}{
   \mathcal{L}_{XY} [\alpha] = |(\partial_\mu - i  \alpha_\mu) \phi|^2 + s |\phi|^2 + u |\phi|^4 \,.
 }
As in the previous examples, we can consider a monopole configuration with $q$ units of background magnetic flux as defining the insertion ${\cal M}_q$.

Unfortunately, the Wilson-Fisher fixed point of a single complex scalar cannot be accessed perturbatively, so we will compute the dimensions $\Delta_q$ by first generalizing $\mathcal{L}_{XY}[\alpha]$ to a theory with $N_b$ complex scalars with Lagrangian
 \es{LGeneral}{
   \mathcal{L} [\alpha] = \sum_{a = 1}^{N_b}  \abs{(\partial_\mu - i  \alpha_\mu) \phi_a} ^2 + s |\vec{\phi} |^2  + u \left( |\vec{\phi} |^2 \right)^2 \,, \qquad
     |\vec{\phi} |^2 \equiv \sum_{a = 1}^{N_b} \abs{\phi_a}^2 \,,
 }
and then performing a $1/N_b$ expansion.  Our goal in this paper is to find the first two terms in this expansion.   The CFT (obtained by setting $\alpha = 0$ in \eqref{LGeneral} and tuning $s$ to zero) is the Wilson-Fisher fixed point with $O(2 N_b)$ symmetry.  The $U(1)$ symmetry that we consider is a subgroup of $O(2N_b)$ that acts by rotating each complex scalar by the same phase.

The rest of this paper is organized as follows.  In Section~\ref{sec:method} we set up our conventions and explain the method we use to compute $\Delta_q$ in the model \eqref{LGeneral} in more detail.  In Section~\ref{INFINITY} we perform the leading order calculation in $N_b$.  To this order, we find agreement with the results of Refs.~\cite{murthy,hermelemono} on the leading large $N_b$ dependence of the dimensions of monopole operators in the $\CP^{N_b-1}$ model.  Indeed, to leading order in $N_b$, one can ignore the contribution to the $S^2$ ground state energy coming from the gauge field fluctuations in the $\CP^{N_b - 1}$ model, so the scale dimensions of the monopole operators in that model should agree with those in the ungauged theory \eqref{LGeneral}.  In Section~\ref{CORRECTIONS} we compute the leading $1/N_b$ corrections to $\Delta_q$.  We end with a discussion of our results in Section~\ref{DISCUSSION}.

\section{Method}
\label{sec:method}

We consider the $O(2N_b)$ scalar field theory defined on an arbitrary conformally flat manifold by the action 
 \es{ActionCurved}{
  {\cal S} = \sum_{a=1}^{N_b} \int d^3 r\, \sqrt{g} \Bigl[ g^{\mu\nu} \left[(\partial_\mu + i \alpha_\mu) \phi_a^\ast \right] \left[ (\partial_\nu - i \alpha_\nu) \phi_a \right] + \left( i \lambda + \frac {\cal R}{8} \right) \abs{\phi_a}^2  \Bigr] \,,
 }
where ${\cal R}$ is the Ricci scalar of the background metric $g_{\mu\nu}$, and $\alpha_\mu$ is a background gauge field.  The only dynamical fields are the complex scalars $\phi_a$ and the Lagrange multiplier field $\lambda$.   
It can be checked explicitly that this action is invariant under the Weyl transformations
 \es{Invariance}{
  g_{\mu\nu} &\to f(r)^2 g_{\mu\nu} \,, \qquad
   \alpha_\mu \to \alpha_\mu \,, \qquad
  \phi_a \to f(r)^{-1/2} \phi_a \,,  \qquad
    \lambda \to f(r)^{-2} \lambda \,,
 }
for which $f$ can be taken to be an arbitrary real-valued function.  We will be interested in the action \eqref{ActionCurved} on two conformally flat backgrounds:  $\R^3$ and $S^2 \times \R$, which have ${\cal R} = 0$ and ${\cal R} = 2$, respectively.

On $\R^3$, in the case where and $\alpha_\mu = 0$, the action \eqref{ActionCurved} describes the Wilson-Fisher fixed point of $2N_b$ real scalars.   Indeed, one can add the term $\int d^3r \, \lambda^2 / (4u)$ to this action without changing the IR fixed point, because $u$ flows to infinity;  integrating out $\lambda$ produces the interacting theory \eqref{LGeneral} with $\alpha_\mu = s = 0$, which represents the more conventional description of the Wilson-Fisher fixed point.  The monopole background $\langle \alpha_\mu \rangle = \mathcal{A}^q_\mu$ that corresponds to an insertion of ${\cal M}_q$ at the origin of $\R^3$ satisfies, in spherical coordinates,\footnote{In standard vector notation, we would write $\vec{\nabla} \times \vec{\mathcal{A}}^q = q \hat{e}_r / (2 \abs{r}^2)$ instead of \eqref{dA}, and $\vec{\cal A}^q = \frac q2 (1 - \cos \theta) / (r \sin \theta) \hat e_\phi$ instead of \eqref{Monopole} in flat space.  On $S^2 \times \R$, we have $\vec{\cal A}^q = \frac q2 (1 - \cos \theta) / (\sin \theta) \hat e_\phi$.}
 \es{dA}{
   d{\cal A}^q = \frac q2 \sin \theta d\theta \wedge d\phi \,, 
 }
which follows from \eqref{qUnits}.  We can work in a gauge where
 \es{Monopole}{
  \mathcal{A}^q(r) = \frac{q}{2} ( 1- \cos \theta) d\phi \,.
 }
This background gauge field is well-defined everywhere away from $\theta = \pi$ where there is a Dirac string.  This Dirac string is not observable provided that $q$ is taken to be an integer.

Starting with the theory on $\R^3$ in the monopole background \eqref{Monopole}, the theory on $S^2 \times \R$ can be obtained from a Weyl transformation as in \eqref{Invariance}.   Indeed, writing the flat metric on $\R^3$ in spherical coordinates as
 \es{R3Metric}{
  ds^2 = dr^2 + r^2 d \Omega^2 \,,
 }
and defining $r = e^{\tau}$, we obtain a metric conformal to $S^2 \times \R$:
 \es{R3MetricConformal}{
   ds^2 = e^{2 \tau} \left( d \tau^2 + d\Omega^2 \right) \,.
 }
So if we send $g_{\mu\nu}^{\R^3} \to e^{-2 \tau} g_{\mu\nu}^{\R^3} = g_{\mu\nu}^{S^2 \times \R}$ and at the same time rescale $\phi_a \rightarrow e^{\tau/2} \phi_a$, $\lambda \to e^{2 \tau} \lambda$, $\alpha_\mu \to \alpha_\mu$ as dictated by \eqref{Invariance}, we obtain the action on $S^2 \times \R$.  The monopole background \eqref{Monopole} now corresponds to a constant magnetic field uniformly distributed over $S^2$.

As explained in the introduction, we identify the scaling dimensions $\Delta_q$ of the monopole insertions ${\cal M}_q$ with the ground state energy ${\cal F}_q$ on $S^2$.    We expand this ground state energy at large $N_b$ as follows:
 \es{expand}{
   \mathcal{F}_q = N_b \mathcal{F}_q^{\infty} + \delta \mathcal{F}_q + {\cal O}(1/N_b) \,.
 }
When $q = 0$ the operator ${\cal M}_q$ is just the identity operator and it corresponds to the ground state on $S^2$ in the absence of any magnetic flux.  We expect this operator to have vanishing scaling dimension.  Indeed, we will check explicitly that ${\cal F}_0 = 0$ in our regularization scheme.

We now turn to the evaluation of ${\cal F}_q^\infty$ in the next section and of $\delta {\cal F}_q$ in Section~\ref{CORRECTIONS}.  We will work solely on $S^2 \times \R$ whose coordinates we denote collectively by $r \equiv (\tau, \theta, \phi)$.

\section{$N_b = \infty$ theory}
\label{INFINITY}

In computing the leading large $N_b$ contribution to the ground state energy on $S^2$, one can evaluate the partition function corresponding to \eqref{ActionCurved} in the saddle point approximation where the fluctuations of the Lagrange multiplier field $\lambda$ can be ignored.  However, $\lambda$ should be adjusted such that the ground state energy is minimized.  We thus expand the Lagrange multiplier about its saddle point value as\footnote{This notation has been chosen to be compatible with Ref.~\cite{hermelemono} that studied the $\CP^{N_b -1}$ model.}
 \es{lambdaExpansion}{
   i \lambda = a_q^2  + \frac{q^2}{4} + i \tilde \lambda \,,
 }
where $a_q^2$ will be determined shortly by the saddle-point condition, and $\tilde \lambda$ is a fluctuation that we will consider in the next section.

We expand the field $\phi_a$ in terms of the monopole harmonics defined in Ref.~\cite{wu}:\footnote{Note that our definition of $q$ differs from that of Ref.~\cite{wu} by a factor of two.}
 \es{zmode}{
  \phi_a (r) = \sum_{\ell=q/2}^{\infty} \sum_{m} 
     \int \frac{d \omega}{2 \pi} Z_{\ell m,a} (\omega) Y_{q/2,\ell m} (\theta, \phi) e^{- i \omega \tau} \,.
 }
The quadratic action for the $\phi_a$ then takes the diagonal form
\beq
\mathcal{S} = \sum_{a = 1}^{N_b} \sum_{\ell = q/2}^{\infty} \sum_{m=-\ell}^{\ell} \int \frac{d \omega}{2 \pi} \left[ \omega^2 + (\ell + 1/2)^2 + a_q^2 \right] 
|Z_{\ell m, a} ( \omega)|^2 \,, \label{Szl} 
\eeq
where we have used the fact that the eigenvalues of the gauge-covariant Laplacian on $S^2$ are $\ell (\ell + 1) - (q/2)^2$ \cite{wu}.  From \eqref{Szl}, it is easy to read off the leading approximation to the ground state energy at large $N_b$, which comes from performing the Gaussian integral over the scalar fields $\phi_a$, or equivalently over the coefficients $Z_{\ell m, a}$.  The coefficient ${\cal F}_q^\infty$ appearing in \eqref{expand} is then \cite{hermelemono}
 \es{FqFirst}{
  {\cal F}_q^\infty =  \int \frac{d \omega}{2 \pi} \sum_{\ell=q/2}^{\infty}
(2 \ell + 1) \log \left[\omega^2 + (\ell + 1/2)^2 + a_q^2 \right] \,.
 }
This expression is divergent, but it can be evaluated, for instance, using zeta function regularization.  First we write formally $\log A = -d A^{-s}/ds \big|_{s=0}$ in all the terms of \eqref{FqFirst}, then we evaluate the sum and integral at values of $s$ where they are absolutely convergent, and at the end we set $s=0$.  Performing the $\omega$ integral, we obtain
 \es{FIntermediate}{
   {\cal F}_q^\infty =  \sum_{\ell=q/2}^{\infty}
(2 \ell + 1)  \left[ (\ell + 1/2)^2 + a_q^2 \right]^{\frac 12 -s }  \Biggr|_{s=0}\,,
 }
which still diverges when evaluated at $s=0$.  We then use the identity
  \es{Divergent}{
     2 \sum_{\ell=q/2}^{\infty} \left[  (\ell+1/2)^{2(1-s)} + \left( \frac 12 - s \right) a_q^2 (\ell + 1/2)^{2s}  \right] \Biggr|_{s=0} =  \frac{q(1-q^2)}{12}  - \frac{q a_q^2}{2}   \,.
  }
This identity can be derived by writing the sums on the left-hand side in terms of the Hurwitz zeta function $\zeta(s, a) = \sum_{n=0}^\infty 1/(n+a)^s$ and analytically continuing to $s = 0$.  The terms on the left-hand side of \eqref{Divergent} are nothing but the large $\ell$ expansion of the terms in \eqref{FIntermediate}, so subtracting \eqref{Divergent} from \eqref{FIntermediate} yields a finite result when $s=0$.  Adding and subtracting \eqref{Divergent} from \eqref{FIntermediate}, we therefore find
 \es{FinfinityFinal}{
    \mathcal{F}_q^{\infty}
        &= 2 \sum_{\ell=q/2}^{\infty} \left[ (\ell + 1/2)  
         \left[(\ell + 1/2)^2 + a_q^2 \right]^{1/2} - (\ell+1/2)^2 - \frac{1}{2} a_q^2   \right] \\
      &\qquad\qquad\qquad{}-2 \left[ \frac{q(q^2-1)}{24}  + \frac{q a_q^2}{4} \right]  \,,
 }
which involves a convergent sum over $\ell$ that can easily be evaluated numerically.

The value of $a_q^2$ is not arbitrary, but should be chosen so that the saddle point condition
 \es{partialF}{
   \frac{\partial \mathcal{F}_q^{\infty}}{\partial a_q^2} = 0 
 }
is satisfied.  In our case, where ${\cal F}_q^\infty$ is given by \eqref{FinfinityFinal}, we therefore have
 \es{solveaq}{
    \sum_{\ell = q/2}^{\infty} \left( \frac{\ell +1/2}{\sqrt{ (\ell + 1/2)^2 + a_q^2}} - 1 \right) = \frac{q}{2} \,.
 }
For the first few small values of $q$, we give in Table~\ref{FLeadingTable} the solutions of this equation as well as the corresponding values of ${\cal F}_q^\infty$ obtained after plugging these solutions back into \eqref{FinfinityFinal}.
The values of $\mathcal{F}_q^{\infty}$ agree precisely with those obtained in Ref.~\cite{murthy} in the large $N_b$ limit of the $\CP^{N_b-1}$ model.
 \begin{table}[ht!]
%\caption{default}
\begin{center}
\begin{tabular}{c|c|c}
  $q$ & $a_q^2$ & ${\cal F}_q^\infty$ \\
  \hline \hline
  $0$ & $0$ & $0$ \\
  $1$ & $-0.4498063$ & $0.1245922$ \\
  $2$ & $-1.3978298$ & $0.3110952$ \\
  $3$ & $-2.8454565$ & $0.5440693$ \\
  $4$ & $-4.7929356$ & $0.8157878$\\
  $5$ & $-7.2403441$ & $1.1214167$
\end{tabular}
\end{center}
\caption{The values of $a_q^2$ that solve \eqref{solveaq} and the corresponding coefficients ${\cal F}_q^\infty$ that enter the large $N_b$ expansion of the ground state energy \eqref{expand} on $S^2$.}
\label{FLeadingTable}
\end{table}%

The $q=0$ case of these results is notable. The value $a_0^2 = 0$ is just that 
expected from the conformal mapping between $\R^3$ and $S^2 \times \R$. Also $\mathcal{F}_0^{\infty} = 0$, 
a result that was not evident at intermediate stages.

\section{$1/N_b$ corrections} 
\label{CORRECTIONS}

\subsection{General structure}

The leading $1/N_b$ correction to the result of the previous sections comes from the contribution to the $S^2$ ground state energy coming from the fluctuations $\tilde \lambda$ of the Lagrange multiplier.   Let us begin by discussing the general structure of this correction.  

After integrating out $\phi_a$, the effective action for the fluctuations takes the form
 \es{EffAction}{
  {\cal S}_\text{eff} &=  
\frac{1}{2} \int d^3\, r d^3 r' \sqrt{g (r) } \sqrt{g (r') }  \tilde \lambda  (r)  D^q (r,r')  \tilde \lambda (r')  + \dotsc \,,
 }
where we omitted higher order terms in $\tilde \lambda$.  The kernel $D^q(r, r')$ appearing in eq.~\eqref{EffAction} is nothing but the two-point correlator of $\abs{\phi_a}^2$ :
 \es{KernelsExplicit}{
  D^q(r, r') &= \langle \abs{\phi_a(r)}^2 \abs{\phi_a(r')}^2 \rangle  = N_b G(r, r') G^*(r, r')  \,, 
  }
where we introduced the Green's function $G(r , r') = \langle \phi^\ast (r) \phi (r') \rangle$ for a single complex field $\phi$ in the background monopole flux ${\cal A}_\mu$.  We will compute this Green's function shortly.  Because of the explicit factor of $N_b$ in \eqref{KernelsExplicit}, at large $N_b$ we can ignore the higher order terms in \eqref{EffAction}, and evaluate the contribution from $\tilde \lambda$ to the partition function in the saddle point approximation.  The coefficient $\delta {\cal F}_q$ appearing in eq.~\eqref{expand} can then be obtained by performing a Gaussian integral, which yields
 \es{deltaF}{
   \delta {\cal F}_q = \frac 12 \log \det D^q \,.
 }

To calculate $\log \det D^q$, we should diagonalize the kernel $D^q$.  This diagonalization is accomplished by expanding $\tilde \lambda$ and $D^q$ in terms of the appropriate spherical harmonics. These quantities do not experience a net monopole flux, because they are neutral, and so we (fortunately) do not need the monopole spherical harmonics here.   The expansions
 \es{lambdaAExpansion}{
  \tilde \lambda(r) &= \int \frac{d\omega}{2 \pi} e^{i \omega \tau} Y_{\ell m}(\theta, \phi) \Lambda_{\ell m}(\omega) \,, \\
   D^q (r,r') &= \int \frac{d \omega}{2 \pi} \sum_{\ell m} D^q_{\ell} (\omega) Y_{\ell m}  (\theta, \phi) Y^\ast_{\ell m} (\theta', \phi')  
      e^{i \omega (\tau - \tau')} 
} 
yield a diagonal effective action
 \es{EffActionFourier}{
  {\cal S}_\text{eff} = \frac 12 \int \frac{d\omega}{2 \pi} \sum_{\ell m}
  D^q_\ell (\omega) |
    \Lambda_{\ell m}(\omega) |^2 \,.
 }
Eq.~\eqref{deltaF} then gives
 \es{FqFinal}{
  \delta \mathcal{F}_q = \frac 12 \int \frac{d\omega}{2 \pi} \sum_{\ell=0}^{\infty}
 ( 2 \ell + 1) \log D^q_\ell (\omega) \,.
 }
In the following subsections we present expressions for the kernel in \eqref{EffActionFourier}:  We first present the simpler kernel at $q=0$, and then the kernels at general $q$.  We will check explicitly that $\delta {\cal F}_0 = 0$, as required by conformal invariance in the absence of any monopole insertions.

\subsection{The kernel of fluctuations at $q=0$}

At $q=0$, it is not hard to obtain the Green's function on $S^2 \times \R$ starting from the Green's function on $\R^3$, $1 / (4 \pi |\vec{r} - \vec{r}'|)$, and using the conformal mapping explained around equation \eqref{R3MetricConformal}.  The result is
 \es{G0}{
   G(r, r') = \frac{1}{4  \pi \sqrt{2( \cosh(\tau-\tau') - \cos \gamma )}} \,,
 }
where $\gamma$ is the relative angle between the two points on $S^2$ defined through
 \es{defgamma}{
    \cos\gamma = \cos\theta \cos \theta^\prime + \sin \theta \sin \theta' \cos(\phi - \phi') \,.
 }

Using \eqref{KernelsExplicit}, \eqref{lambdaAExpansion}, and \eqref{G0}, we obtain
 \es{d0l}{
    D_\ell^0 (\omega) &= \frac{1}{16 \pi} \int_{-\infty}^{\infty} d  \tau \int_0^{\pi} \sin \theta d \theta
      \frac{ e^{i \omega \tau} P_\ell (\cos\theta)}{( \cosh\tau - \cos \theta )} \,.
 }
We performed these integrals analytically for a number of small values of $\ell$; from the structure of these answers we deduced the general result:
 \es{Dell0Gamma}{
  D_\ell^0(\omega) = \abs{\frac{\Gamma\left( \left( \ell+1 + i \omega \right)/2 \right)}{4 \Gamma\left( \left( \ell+2 + i \omega \right)/2 \right)}}^2 \,,
 }
which can be written more explicitly as
 \es{dexact}{
   D_{2\ell}^0 ( \omega)  &= \left[\frac{\tanh(\pi \omega/2)}{8 \omega} \right] \prod_{n=1}^{\ell} \frac{(\omega^2 + (2n-1)^2)}{(\omega^2 + 4n^2)} \,, \\
   D_{2\ell+1}^0 ( \omega)  &= \left[\frac{\omega \coth(\pi \omega/2)}{8(\omega^2 + 1)} \right] \prod_{n=1}^{\ell} \frac{(\omega^2 + 4n^2)}{(\omega^2 + (2n+1)^2)} \,.
 }

In the limit of large $\omega$ and $\ell$ we expect the $\tilde \lambda$ self-energy to be given by the flat space limit
 \es{SelfEnergy}{
   \int \frac{d^3 p'}{8 \pi^3} \frac{1}{p^{'2} (p+ p')^2} = \frac{1}{8 p} \,.
 }
Indeed, expanding \eqref{Dell0Gamma} with the help of the Stirling approximation, we find
 \es{Dell0Expansion}{
  D_\ell^0(\omega) = \frac{1}{8 \sqrt{\omega^2 + \ell(\ell+1)}} - \frac{\ell(\ell+1)}{32 \left(\omega^2 + \ell(\ell+1) \right)^{5/2}} + {\cal O} \left(\frac{1}{\left(\omega^2 + \ell(\ell+1) \right)^{5/2}} \right) \,,
 }
which agrees with \eqref{SelfEnergy} upon using the identification $p \sim \sqrt{\omega^2 + \ell(\ell+1)}$.

\subsection{The kernel of fluctuations for general $q$}

Now we turn to the much harder case of non-vanishing $q$.  In this case we don't have a simple closed form expression for the scalar Green's function, so we turn to the mode expansion \eqref{zmode}.  From the action \eqref{Szl} we deduce that the Green's function for a single $\phi_a$ is
 \es{Gq}{
G (r, r') &= \sum_{\ell = q/2}^{\infty} \int \frac{d \omega}{2 \pi} e^{i \omega(\tau - \tau')} \left[\sum_{m=-\ell}^\ell Y_{q/2,\ell m}^\ast (\theta, \phi)
Y_{q/2,\ell m} (\theta', \phi') \right] \frac{1}{\omega^2 + (\ell + 1/2)^2 + a_q^2} \\
&= \sum_{\ell = q/2}^{\infty}  e^{i q \Theta} F_{q,\ell} ( \gamma) \frac{e^{- E_{q\ell} |\tau - \tau'|}}{2 E_{q\ell}} \,.
 }
In writing the second line we defined the energy
 \es{EDef}{
   E_{q\ell} \equiv \sqrt{ (\ell+1/2)^2 + a_q^2} \,,
 }
and performed the $\omega$ integral;  we also performed the sum over $m$, which, up to the phase factor
\beq
e^{i \Theta}  =\frac{1}{ \cos (\gamma /2)} \left [ \cos (\theta/2) \cos(\theta'/2) + e^{-i (\phi - \phi')} \sin (\theta/2) \sin(\theta'/2) \right ]
\label{defTheta}
\eeq
discussed in Ref.~\cite{wu2}, yields a polynomial in $\cos \gamma$ that can also be written in terms of the monopole harmonics as
 \es{FDef}{
  F_{q,\ell} (\gamma) \equiv \sqrt{ \frac{2 \ell + 1}{4 \pi}}Y_{q/2,\ell,-q/2} (\gamma,0)  \,.
 }
(See Appendix~\ref{MONOPOLEHARMONICS} for more explicit expressions for $F_{q, \ell}(\gamma)$.) Here, $\gamma$ is the relative angle of the two points on $S^2$ defined in \eqref{defgamma}.

From \eqref{KernelsExplicit} and \eqref{Gq}, we can now determine $D^q(r, r')$.  Further extracting $D^q_\ell (\omega)$ using \eqref{lambdaAExpansion} we obtain 
\bea
D^q_\ell (\omega) (2 \pi) \delta (\omega + \omega') &=& \frac{1}{(2 \ell + 1)} \sum_{\ell', \ell''=q/2}^{\infty} \int d^3 r d^3 r' \sqrt{g (r)} \sqrt{g(r')}  
F_{0,\ell} (\gamma) F_{q,\ell'} (\gamma) F_{q,\ell''} (\gamma)  \nn
&~&~~~~~~~~~~~~~ \times \frac{ e^{-(E_{q \ell'}  + E_{q \ell''})|\tau - \tau'| - i \omega \tau - i \omega' \tau'}}{4 E_{q\ell} E_{q \ell'}}  \,.
\eea
We can simplify this expression to
 \es{GotDq}{
D^q_\ell ( \omega ) &= \frac{8 \pi^2}{(2 \ell + 1)} \sum_{\ell' \ell''=q/2}^{\infty}\ \left[ \frac{ E_{q\ell'}  + E_{q \ell''} }
{
2  E_{q\ell'}  E_{q\ell''} ( \omega^2 + ( E_{q\ell'} +  E_{q\ell''})^2)
}
\right] \mathcal{I}_D (\ell,\ell',\ell'') \,,
 }
where
\beq 
\mathcal{I}_{D} (\ell, \ell', \ell'') = \int_0^\pi \sin \theta d \theta F_{0,\ell} (\theta) F_{q,\ell'} (\theta) F_{q,\ell''} (\theta) \,.
\eeq
This is an integral of three monopole harmonics and can be expressed in terms of the Wigner 3-$j$ symbols as
\beq
\mathcal{I}_D (\ell, \ell', \ell'') = \left[ \frac{(2 \ell + 1) (2 \ell' + 1) (2 \ell'' + 1)}{32 \pi^3} \right]
\left( 
\begin{array}{ccc} \ell & \ell' & \ell'' \\ 0 & -q/2 & q/2 \end{array} \right)^2  \,.
\label{ID}
\eeq
We can check, for instance, that this result equals (\ref{d0l}) for $q=0$ and $\ell=0$
\bea
D_0^0 (\omega) &=& \frac{1}{2 \pi} \sum_{\ell = 0}^{\infty} \frac{1}{\omega^2 + (2 \ell + 1)^2} = \frac{ \tanh(\pi \omega/2)}{8 \omega} \,.
\eea

\subsection{Numerics}

The results of the previous sections are all we need for calculating numerically the correction $\delta {\cal F}_q$ to the scaling dimensions of the monopole operators.  Unfortunately, the expression \eqref{FqFinal} is formally divergent, as can be seen for instance in the case $q=0$ where we know $D^0_\ell(\omega)$ explicitly, and hence eq.~\eqref{FqFinal} is not suitable for numerical evaluation in its current form.  However, we expect the divergences to be independent of $q$, so the differences $\delta {\cal F}_{q_1} - \delta {\cal F}_{q_2}$ should be finite and shouldn't require regularization.  Moreover, it must be true that $\delta {\cal F}_0 = 0$, because the case $q=0$ corresponds to an insertion of the identity operator, which should have vanishing scaling dimension.  (See Appendix~\ref{DELTAF0} for an explicit check that $\delta {\cal F}_0 = 0$.)  Subtracting $\delta {\cal F}_0$ from \eqref{FqFinal}, we can then also write $\delta {\cal F}_q$ as
 \es{FqFinalReg}{
\delta \mathcal{F}_q = \frac 12 \int \frac{d\omega}{2 \pi} \sum_{\ell=1}^{\infty}
 ( 2 \ell + 1) \log \frac{D^q_\ell (\omega)}{D^0_\ell (\omega)} \,,
 }
which we evaluate numerically for the first few lowest values of $q$.

In evaluating \eqref{FqFinalReg}, one has to perform three sums (two when calculating $D_\ell^q(\omega)$ using \eqref{GotDq}  and one in \eqref{FqFinalReg}) and one integral over $\omega$.  Let us first comment on the two sums in \eqref{GotDq}.  For fixed $\ell'$, the sum over $\ell''$ in \eqref{GotDq} has only finitely many non-zero terms because the $3$-$j$ symbols in \eqref{ID} vanish unless $\ell$, $\ell'$ and $\ell''$ satisfy the triangle inequality.  To see whether or not the remaining sum over $\ell'$ is convergent, one should find an asymptotic expansion at large $\ell'$ for the terms in this sum.  While for general $\ell$ it may seem hard to do so, it is easier to first fix $\ell$ to a small value for which the sum over $\ell''$ has $2\ell+1$ terms that can be written down explicitly, and the large $\ell'$ asymptotics can be easily computed.  Repeating this procedure for several values of $\ell$, one can infer the large $\ell'$ asymptotics for all $\ell$ by noticing that all the expressions involved are polynomials in $\ell(\ell+1)$.  The first few terms are
 \es{ellpAsymp}{
  \frac{1}{8 \pi \ell'^2} - \frac{1}{8 \pi \ell'^3} + \frac{3 - 6 a_q^2 + 2 \ell(\ell+1) - \omega^2}{32 \pi \ell'^4} + \ldots \,.
 }
This expression shows that the sum over $\ell'$ is absolutely convergent.  To save computational resources, one can use a mix of numerical and analytical techniques in evaluating $D_\ell^q(\omega)$:  the terms with low $\ell'$ should be summed up explicitly, while for the terms with large $\ell'$ one can sum up analytically the approximate expression \eqref{ellpAsymp} developed to a higher order of accuracy. (In our computations, we developed the large $\ell'$ approximation up to order $1/\ell'^{13}$.)

\begin {figure} [tb]
  \center\includegraphics [width=0.6\textwidth] {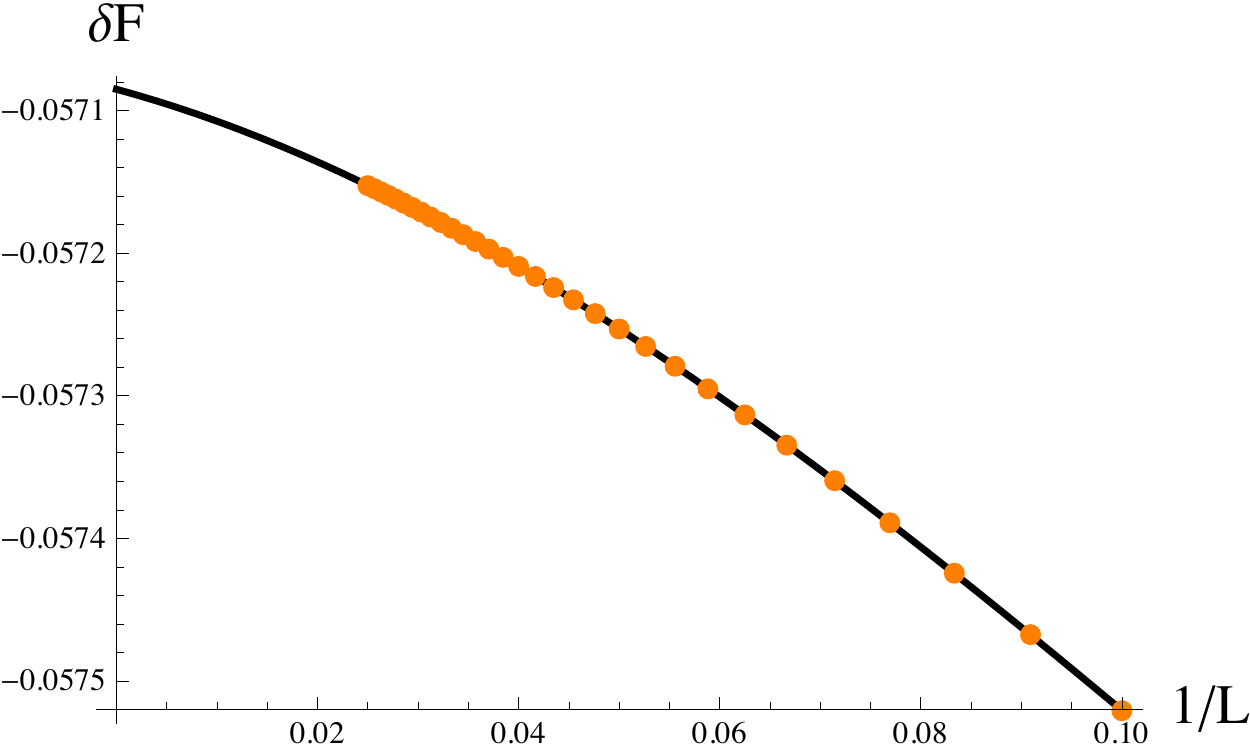}
  \caption {The coefficient $\delta {\cal F}_1$ evaluated numerically from \eqref{FqFinalReg} using the relativistic cutoff \eqref{Relativistic} as a function of the inverse cutoff scale $1/L$.  The solid line is a quadratic fit from which we extract the value $\delta {\cal F}_1 \approx -0.057$ as we take $L \to \infty$.\label {FPlot}}
\end {figure}

Lastly, in calculating \eqref{FqFinalReg} one should be wary that there could still be divergences.  We find that imposing a relativistic cutoff\footnote{At high energies the Lorentzian theory has $SO(2,1)$ symmetry that is also obeyed by the cutoff \eqref{Relativistic}, so the speed of light is not renormalized. If one chooses a cutoff that breaks the $SO(2,1)$ symmetry, then there are finite corrections to the speed of light in the IR that have to be accounted for.}
 \es{Relativistic}{
  \omega^2 + \ell(\ell+1) \leq L(L+1) \,,
 }
yields a finite answer as we take $L \to \infty$.  The absence of divergences relies heavily not only on the choice of cutoff \eqref{Relativistic}, but also on choosing the value of $a_q^2$ that solves eq.~\eqref{solveaq};  for other values of $a_q^2$ there would be divergences.   See Figure~\ref{FPlot} for a plot of $\delta {\cal F}_q$ in terms of $1/L$ in the case $q = 1$, where from the large $L$ extrapolation we obtain $\delta {\cal F}_1 \approx -0.057$.
In this case we therefore conclude that the scaling dimension of the monopole operator ${\cal M}_1$ is
 \es{Delta1}{
  \Delta_1 = 0.125 N_b -0.057 + {\cal O}(1/N_b) \,,
 }
where we included the leading large $N_b$ behavior that was also given in Table~\ref{FLeadingTable}.  Repeating this procedure for the first few small values of $q$, we obtain the results in Table~\ref{FCombinedTable}.
 \begin{table}[ht!]
%\caption{default}
\begin{center}
\begin{tabular}{c|c}
  $q$ & $\Delta_q = {\cal F}_q$ \\
  \hline \hline
  $0$ & $0$ \\
  $1$ & $0.125 N_b -0.057 + {\cal O}(1/N_b)$ \\
  $2$ & $0.311 N_b -0.152 + {\cal O}(1/N_b)$ \\
  $3$ & $0.544 N_b -0.272 + {\cal O}(1/N_b)$ \\
  $4$ & $0.816 N_b -0.414 + {\cal O}(1/N_b)$\\
  $5$ & $1.121 N_b -0.575 + {\cal O}(1/N_b)$
\end{tabular}
\end{center}
\caption{The scaling dimensions of the first few monopole operators ${\cal M}_q$ in the Wilson-Fisher CFT of $N_b$ complex scalars in the large $N_b$ expansion \eqref{expand}.  The leading large $N_b$ behavior was computed in Section~\ref{INFINITY}, and agrees with results from the $\CP^{N_b-1}$ model \cite{murthy}.  The ${\cal O}(N_b^0)$ term was computed numerically using \eqref{FqFinalReg}.}
\label{FCombinedTable}
\end{table}%
This is the main result of this paper.

\section{Discussion}
\label{DISCUSSION}

Following recent work \cite{willett,dopedcft}, in this paper we considered monopole insertions in $2+1$-dimensional CFTs that have a global $U(1)$ symmetry.
A simple example of such a CFT is the Wilson-Fisher fixed point of the $XY$ model.  Critical exponents of this CFT have long been the focus of much study, and are among the most accurately known non-trivial exponents of higher dimensional CFTs \cite{vicari}.
Associated with the monopole  insertions, we have a new set of critical exponents of this venerable CFT\@.  We computed these exponents ({\em i.e.\/}~monopole scaling dimensions) to next-to-leading order in the $1/N_b$ expansion of a theory with $N_b$ complex bosons.
Our results for the scaling dimensions are summarized in Table~\ref{FCombinedTable}.

The numerical series in Table~\ref{FCombinedTable} appear to be reasonable even when evaluated at $N_b=1$.
It would be interesting to also compute the monopole scaling dimensions in Monte Carlo simulations or series expansions, 
such as those in Ref.~\cite{vicari}.

\section*{Acknowledgments}

We thank E.~Dyer, M.~Headrick, A.~Kapustin, M.~Mezei, and D.~Neill for useful discussions.  The work of SSP is supported in part by a Pappalardo Fellowship in Physics at MIT and in part by the U.S. Department of Energy under cooperative research agreement Contract Number DE-FG02-05ER41360\@.  The work of SS was supported by the U.S.\ National Science Foundation under grant DMR-1103860 and by the U.S.\ Army Research Office Award W911NF-12-1-0227.

\appendix

\section{Free scalar theory}
\label{SCALAR}

We can also calculate the scaling dimensions $\Delta_q$ in the free theory of $N_b$ complex scalars.  The only difference from the Wilson-Fisher CFT is that the action for the free theory does not have a Lagrange multiplier $\lambda$, but there is a conformal coupling ${\cal R} \abs{\phi}^2$ in the action, as in \eqref{ActionCurved}.  The ground state energy on $S^2$ in the presence of $q$ units of magnetic flux that we obtain by integrating out the scalars is $N_b {\cal F}_q^\infty$, where ${\cal F}_q^\infty$ can be computed from \eqref{FinfinityFinal} with $a_q^2 = -q^2/4$, as appropriate for conformally coupled scalars.  See Table~\ref{FFreeTable} for a few particular cases.  These results are exact.
 \begin{table}[ht!]
%\caption{default}
\begin{center}
\begin{tabular}{c|c}
  $q$ & $\Delta_q / N_b$ \\
  \hline \hline
  $0$ & $0$ \\
  $1$ & $0.097$ \\
  $2$ & $0.226$ \\
  $3$ & $0.384$ \\
  $4$ & $0.567$\\
  $5$ & $0.770$
\end{tabular}
\end{center}
\caption{The first few scaling dimensions $\Delta_q$ of the monopole insertions ${\cal M}_q$ in the free CFT of $N_b$ scalars.}
\label{FFreeTable}
\end{table}%

\section{Monopole harmonics}
\label{MONOPOLEHARMONICS}

We start with the relation
 \es{Ysum}{
    \sum_{m=-\ell}^{\ell} Y^\ast_{q/2,\ell m} (\theta, \phi) Y_{q/2,\ell m} (\theta', \phi') =  F_{q,\ell} (\gamma) e^{iq \Theta}  \,, 
 }
where $F$ is defined in eq.~(\ref{FDef}) and the angles $\gamma$ and $\Theta$ are defined in eq.~(\ref{defgamma}).

Above, we have used the functions
 \es{FExpressions}{
F_{q,\ell} (\theta) &\equiv  \sqrt{\frac{(2 \ell + 1)}{4 \pi}} Y_{q/2,\ell,-q/2} (\theta,0) \\
&= 2^{-q/2} \left(\frac{2 \ell+ 1}{4 \pi}\right) (1 + \cos\theta)^{q/2} P^{0,q}_{\ell - q/2} (\cos \theta) \\
&= 2^{-q/2} \left(\frac{2 \ell+ 1}{4 \pi}\right) (1 + \cos\theta)^{q/2-1} \left[ \frac{ (\ell + q/2) P_{\ell-q/2}^{0,q-1} (\cos \theta) + (\ell - q/2 + 1) P_{\ell - q/2 + 1}^{0,q-1} (\cos\theta)}{(\ell + 1/2)} \right] \,.
 }
The special values are
 \es{FLegendre}{
   F_{q,\ell} (\theta) = \begin{cases} 
\displaystyle \left(\frac{2 \ell+ 1}{4 \pi}\right) P_\ell (\cos \theta) &  \text{if $q=0$} \,, \\
\displaystyle \frac{1}{\sqrt{2}}\left(\frac{2 \ell+ 1}{4 \pi}\right) (1+\cos\theta)^{-1/2}\left[P_{\ell-1/2} (\cos \theta) + P_{\ell + 1/2} (\cos \theta) \right] & \text{if $q=1$} \,,
  \end{cases}
 }
etc.

\section{Calculation of $\delta {\cal F}_0$}
 \label{DELTAF0}

We now show that using zeta-function regularization we find $\delta {\cal F}_0 = 0$.  Using the infinite product representation for the hyperbolic tangent and cotangent in \eqref{dexact}, one can show that
 \es{logD}{
  \log D_\ell^0(\omega) = \sum_{k=\ell+1}^\infty (-1)^{k+\ell} \log (\omega^2 + k^2 ) + \text{($\omega$-independent terms)} \,.
 }
The $\omega$-independent terms do not contribute to $\delta {\cal F}_0$ in our regularization scheme.  With the help of
 \es{intomega}{
  \int \frac{d\omega}{2 \pi} \log (\omega^2 + a^2) = \abs{a} \,,
 }
which can be derived, for instance, by rewriting \eqref{intomega} as
 \es{intomega2}{
  -\frac d{ds} \int \frac{d\omega}{2 \pi} \frac{1}{\left(\omega^2 + a^2\right)^s} \Biggr|_{s=0} = -\frac d{ds} \frac{\sqrt{\pi} \Gamma(s - 1/2) \abs{a}^{1 - 2s}}{2 \pi \Gamma(s)} \Biggr|_{s=0} = \abs{a} \,,
 }
we can perform the $\omega$ integral in \eqref{FqFinal} and we obtain
 \es{deltaF0Sum}{
  \delta {\cal F}_0 = \frac 12 \sum_{\ell=0}^\infty (-1)^\ell (2 \ell+1) \sum_{k=\ell+1}^\infty (-1)^k k \,.
 }
The sum over $k$ can be written in terms of the Hurwitz zeta function $\zeta(s, a) = \sum_{n=0}^\infty 1/(n+a)^s$ as
 \es{kSum}{
  \sum_{k=\ell+1}^\infty (-1)^k k = 2 (-1)^\ell \left[\zeta \left(-1, \frac{\ell+2}{2} \right) - \zeta \left( -1, \frac{\ell+1}{2} \right)  \right]
   = \frac{(-1)^{\ell+1}}{4} (2\ell+1) \,,
 }
so then
 \es{deltaF0Zero}{
  \delta {\cal F}_0 = -\frac 12 \sum_{\ell=0}^\infty \left(\ell+\frac 12 \right)^2  = -\frac 12 \zeta \left(-2, \frac 12 \right) = 0\,.
 }

\end{document}